\documentclass[twocolumn]{article}

\usepackage{makecell}
\usepackage{graphicx}
\usepackage{amsmath}
\usepackage{booktabs}
\usepackage{array}
\usepackage{authblk}
\usepackage{algorithm}
\usepackage{algpseudocode}
\floatname{algorithm}{Algorithm}
\usepackage{caption}
\usepackage[numbers,sort&compress]{natbib}
\usepackage[hidelinks]{hyperref}
\usepackage{titlesec}
\usepackage{amsmath}

\title{A Principled Framework for Safe Algorithm Updates in Automated Insulin
Delivery Systems}

\author[1]{Thomas Screven\thanks{tsscreven@ucdavis.edu}}
\author[1]{Ziqiang ``Joe'' Zhu\thanks{ziqzhu@ucdavis.edu}}
\author[2]{Deniz Cengiz\thanks{d.c.cengiz@nightscoutfoundation.org}}
\author[3]{Rayhan A. Lal\thanks{inforay@stanford.edu}}
\author[3]{Korey K. Hood\thanks{kkhood@stanford.edu}}
\author[1]{Samuel T. King\thanks{kingst@ucdavis.edu}}
\affil[1]{University of California, Davis}
\affil[2]{Trio OS-AID}
\affil[3]{Stanford University}

\date{}

\begin{document}

\maketitle

\begin{abstract}

\textbf{Background:} AID algorithms require ongoing software updates and
bug fixes. However, in co-adapted systems --- where users tune settings
around existing algorithmic behavior --- fixing bugs can paradoxically
disrupt glycemic control. No principled framework exists to evaluate the
safety of AID algorithm updates.

\textbf{Methods:} We developed a two-part framework to classify bugs and
evaluate the \emph{clinical equivalence} of software updates in AID
systems. Bugs are classified as factual, heuristic, or computational,
each with distinct management strategies. These classifications were
validated during a port of Trio's \texttt{oref} algorithm from Javascript to a
bug-fixed Swift implementation. We compared implementations using shadow
execution on 736,480 invocations from eight Trio users. The second
component assesses clinical equivalence through error analysis on paired
glucose values, applied to both Trio implementations using mechanistic
in silico simulation and data-driven replay simulation.

\textbf{Results:} In mechanistic in silico simulation, the Swift
and Javascript implementations produced nearly identical Time in Range
(84.9\% vs.~84.9\%) and Glycemia Risk Index (23.5\% vs.~23.9\%), with
$\geq$99\% of paired glucose values in Parkes Error Grid Zones A and B,
meeting our threshold for clinical equivalence. Shadow execution showed
low mismatch rates in \texttt{oref} components (\texttt{iob} 0.43\%, \texttt{autosens} 1.22\%,
\texttt{determineBasal} 0.07\%, \texttt{meal} 0.01\%), with clinically meaningful
differences in 0.03\% of \texttt{iob} invocations. Data-driven replay simulations
of bugs revealed $\geq$99\% of downstream paired glucose values in Parkes
Error Grid Zones A and B, also meeting our threshold for clinical
equivalence.

\textbf{Conclusions:} Our framework integrates bug-fixing principles
with multi-method clinical evaluation to assess the safety of AID
algorithm updates. It is system-agnostic and applicable to all widely
used OS-AID systems, with case studies highlighting the need for
systematic remediation of factual and computational bugs.

\end{abstract}


\section{Introduction}\label{introduction}

Automated insulin delivery (AID) systems are used by thousands of people
with diabetes every day. Unlike ordinary software, some AID systems that
allow user-adjustable settings create a co-adaptation loop: users tune
therapy parameters over weeks and months in response to algorithmic
behavior \cite{kovatchev25, messer25}. The system's effective behavior
is a joint product of the code and the human response to it.

This creates a fundamental problem for software maintenance. When the
underlying algorithm changes, even correctly, it can break the
co-adapted equilibrium and paradoxically worsen glycemic control
\cite{kovatchev25}. If an algorithm underestimates insulin impact, users
compensate by lowering sensitivity settings; after the bug is fixed,
those same settings cause overdosing. In traditional software, bugs are
bad and fixes are good. AID systems break this model.

Yet AID systems require ongoing updates: bug fixes, platform ports, and
algorithm improvements. Each change carries potential clinical
consequences for the people who depend on it. Currently, no principled
framework exists for evaluating the safety of algorithmic and other
software updates to AID systems.

To address this, we developed a two-part framework that classifies and
evaluates AID algorithm updates. The framework's goal is not absolute
safety but \emph{clinical equivalence}: demonstrating that changes to a
validated algorithm produce quantitatively similar behavior under
realistic conditions.

Part I classifies patches into three categories: factual (objective
physiological computations that should be corrected immediately),
heuristic (statistical or adaptive estimates that should be preserved or
transitioned with monitoring), and computational (numerical precision
differences).

Part II quantifies clinical equivalence for algorithm changes using
three complementary methods: in silico simulation to test end-to-end
algorithm behavior, shadow execution to surface differences between
patched and original versions on real-world data, and data-driven
replay.

We validated this framework in the context of Trio, an open-source AID
(OS-AID) system, where we identified bugs spanning logic errors,
floating-point precision differences, input mutation side effects, and
heuristic threshold errors while porting the \texttt{oref} algorithm from
Javascript to Swift. We also found that \texttt{oref} is highly discontinuous
where small numerical differences cascade through branching logic into
large output divergences, making the impact of any change difficult to
predict by inspection alone.

We chose Trio as the empirical subject for this framework because it
sits at the intersection of clinical maturity and active software
evolution. OS-AID systems are no longer experimental: a landmark
randomized controlled trial demonstrated that open-source algorithms are
as safe and effective as commercial systems \cite{burnside22},
prospective observational studies have confirmed their real-world safety
at scale \cite{lum21}, and an international consensus statement now
formally recognizes their clinical role \cite{braune22}. This clinical
legitimacy raises the stakes for software maintenance --- Trio's users
are not research participants but individuals whose glycemic control
depends on the algorithm's stability across updates. At the same time,
Trio is under continuous development, making it an ideal testbed for a
framework designed to evaluate exactly this kind of ongoing change. We
release the Swift \texttt{oref} implementation \cite{triodev} and shadow
execution replay tooling for the broader OS-AID community.

\section{Decision framework}\label{decision-framework}

In our decision framework we classify AID algorithm bugs into three
categories based on the nature of the computation, not the severity of
the bug. Each category has a distinct management strategy. We also
define \emph{clinical equivalence} to quantify similarity between
algorithm versions.

\textbf{Factual components} compute objective physiological quantities.
The primary example is insulin on board (IOB), an accounting metric that
tracks how much insulin is still active in the body at a given time. IOB
has a clear mathematical definition when it is used in an AID system: it
is the sum of all recent insulin deliveries, each decayed using a model
of insulin absorption \cite{hovorka04, zisser08}. When the IOB
implementation is wrong, the system is making dosing decisions based on
incorrect facts. Even if users have unknowingly compensated for an IOB
error by adjusting their settings, this compensation is fragile.
Incorrect facts impact everything, making reasoning about an incorrect
factual algorithm nearly impossible.

\emph{We fix factual bugs because these facts serve as the foundation
upon which the entire algorithm is built.}

\textbf{Heuristic components} make statistical or adaptive estimates
where no single correct answer exists. The primary examples are
algorithms to estimate insulin sensitivity, meal absorption modeling,
and glucose forecasting. All of these algorithms use the individual's
current data and statistical methods to approximate the current or
future state, allowing the \texttt{oref} algorithm to adapt to physiological
changes and make predictive dosing decisions. Because these heuristic
modules are inherently approximate and deeply co-adapted with user
behavior over time, fixing a bug can paradoxically disrupt glycemic
control \cite{lewis19, kovatchev25}. For example, a sudden mathematical
``correction'' that shifts an \texttt{autosens} ratio from 1.0 to 0.5 effectively
doubles insulin dosing.

\emph{Thus, we intentionally preserve legacy behaviors, including
confirmed bugs, when porting these components, explicitly staging any
corrections to be transitioned slowly alongside simulation and user
monitoring.}

\textbf{Computational components} are logically equivalent across
implementations but produce different outputs due to using precise and
slow vs.~approximate and fast mathematical operations. In our case, this
means the difference between Javascript's IEEE 754 double-precision
floating-point arithmetic (approximate and fast) and Swift's
arbitrary-precision Decimal type (precise and slow). These differences
are not logic errors: the two implementations encode the same algorithms
using the same operations. But small precision differences can cascade
through branching logic into large output differences.

\begin{algorithm}[h!]
\caption{Set Temp Basal}
\label{alg:set_temp_basal}
\begin{algorithmic}[1]
\Require current basal rate $r_{\text{curr}}$
\State $r_{\text{new}} \gets \textsc{round\_basal}(r_{\text{curr}})$
\State $r_{\text{upper}} \gets r_{\text{curr}} \times 1.2$
\State $r_{\text{lower}} \gets r_{\text{curr}} \times 0.8$
\If{$r_{\text{lower}} \leq r_{\text{new}} \leq r_{\text{upper}}$}
    \State \Return $r_{\text{curr}}$ \Comment{Don't change basal rate}
\Else
    \State \Return $r_{\text{new}}$ \Comment{Program new basal rate}
\EndIf
\end{algorithmic}
\end{algorithm}

Algorithm~\ref{alg:set_temp_basal} is from one part of the Javascript implementation where we
observed mismatches. Imprecision in the bounds calculations or the rate
used by the rounding function can lead to Trio programming different
basal rates on the pump. We observed differences of 20\% or more despite
the calculations and values all being mathematically identical.

\emph{We use precise calculations by default for all algorithms. We fall
back to approximate floating point only in key areas where there is a
measurable performance degradation and the result flows through decision
boundaries that use rounding operations to minimize differences.}

We define \textbf{clinical equivalence} as the property that an updated
algorithm produces glycemic behavior indistinguishable from a clinically
validated reference algorithm within pre-specified margins, evaluated
through in silico methodology. Clinical equivalence is not a claim of
absolute clinical safety. Instead, when equivalence holds, the updated
algorithm inherits the safety profile already established for the
reference through prior clinical validation.

We operationalize clinical equivalence with two pre-specified margins.
The primary margin uses Parkes Error Analysis \cite{parkes00}. Although
originally designed to quantify new glucose meter technology, Parkes
Error Analysis provides a way to compare two glucose values with
validated zones that categorize differences in a clinically meaningful
way for people with diabetes. We consider two glucose traces to be
clinically equivalent if $\geq 99\%$ of glucose values are within Zones A and
B \cite{kuroda17}.

The secondary margin is Glycemia Risk Index (GRI) \cite{klonoff23},
where clinical equivalence is $\Delta GRI \leq 10$. No established precedent exists
for GRI-based equivalence thresholds; our threshold is motivated by the
zone structure of the original GRI framework, in which clinically
distinct risk categories span 20-point intervals.

We additionally report Time in Range (TIR) as a standard descriptive
metric for glycemic control, without a pre-specified equivalence margin.

We applied this framework to the Trio port as follows. The \texttt{oref}
algorithm consists of four major modules: an IOB calculation (\texttt{iob}), a
carbohydrate absorption algorithm (\texttt{meal}), an insulin sensitivity
estimator (\texttt{autosens}), and a dosing logic module that integrates the
outputs of the other three functions, using glucose forecasting, to
compute insulin dosing (\texttt{determineBasal}). The \texttt{iob} algorithm is factual
--- we fixed all logic bugs we identified. The \texttt{meal}, \texttt{autosens}, and
\texttt{determineBasal} algorithms are heuristic --- we preserved existing
behavior except for a small number of egregious errors. We implemented
all functions with precise arithmetic in Swift except for the \texttt{iob}
pharmacokinetic curve calculation, which uses floating-point arithmetic
for performance. Its outputs flow through decision boundaries that use
rounding, which absorbs precision differences.

When applying this framework to Trio, there are ambiguities. In
particular, when a bug in a factual component (like \texttt{iob}) produces
divergent outputs in a heuristic component that consumes it (like
\texttt{autosens}), we must decide whether to fix it. In our framework, we
classify bugs by the component where the error originates, not by the
components where its effects are observed. Thus, bugs in factual
calculations are fixed even when the bulk of their observable impact
appears in heuristic components (see the Discussion section).

\section{Empirical validation}\label{empirical-validation}

Our empirical validation uses two different types of evaluation: shadow
execution to capture inputs from mismatch algorithm runs for evaluating
individual algorithm invocations and simulation for evaluating longer
algorithm runs. Our goal is to compare two algorithms directly and to
assess their equivalence given our changes when porting from Javascript
to Swift.

\subsection{Mechanistic in silico simulation}\label{mechanistic-in-silico-simulation}

To demonstrate the impact of changes in an end-to-end closed-loop
system, we conducted mechanistic in silico simulation
experiments. Our experiments used \texttt{simglucose} \cite{simglucose}, an
open-source implementation of the UVA/Padova Type 1 Diabetes Simulator.
We evaluated both the original Javascript and updated Swift algorithms
across the full 30-patient in silico cohort (10 adults, 10
adolescents, and 10 children) to test against highly diverse, non-linear
physiological profiles.

For each virtual patient, we enabled super-micro-bolus (SMB) and
unannounced meal (UAM) dosing so that the algorithm handles meals
automatically without user announcement. We initialized therapy settings
from each patient's \texttt{simglucose} defaults, which were too aggressive for
several virtual patients in our initial runs. To establish a co-adapted
baseline, we ran a separate tuning scenario against the original
Javascript algorithm. For each virtual patient, we simulated 14 days of
random meal scenarios. For patients whose default settings produced time
below range (\textless70 mg/dL) above the ADA-recommended 4\% threshold
\cite{battelino19}, we increased insulin sensitivity until they fell
under this threshold; patients already meeting the threshold retained
their default settings. For two virtual patients, representing children
aged 9 and 10, we reached a plateau where time below range remained
above 4\% and stopped meaningfully decreasing despite increases in
insulin sensitivity. However, time above range began increasing
significantly due to the virtual patient receiving too little insulin.
For these two patients, we stopped tuning their insulin sensitivity when
this occurred.

After completing insulin sensitivity adjustment, we held these tuned
settings fixed and evaluated the original Javascript algorithm against
the updated Swift algorithm using a custom evaluation scenario, directly
testing whether the ported algorithm preserves glycemic control under
co-adapted settings. Our evaluation scenario used a standard three-meal
input. Meals were predefined to occur within a 60 minute range. Each
minute within a meal window had an equal chance of being chosen as the
meal time. In addition, there was a 50\% chance that a snack occurred
once per day. The timing of a snack was calculated as the midpoint
between breakfast and lunch or lunch and dinner, with both candidate
snack times having an equal chance of being selected. Meal carbohydrate
amounts ranged from 27g to 33g for breakfast, 54g to 66g for lunch, 45g
to 55g for dinner, and 18g to 22g for a snack. For each virtual patient,
both algorithm variants were evaluated on matched scenarios with
identical meal schedules and CGM noise seed.

\subsection{Shadow execution}\label{shadow-execution}

To measure the impact of our bug fixes and precision differences, we ran
shadow execution on eight Trio users from March 2025 - December 2025. In
live systems, both the original Javascript algorithm and the updated
Swift algorithm ran on identical inputs at each algorithm invocation.
When the outputs differed beyond the approximate matching thresholds
defined in the Trio test suite \cite{triotest}, we recorded the full
input state for later replay and analysis.

During replay and analysis, we used different implementations to
understand the impact of different changes. These implementations
included the unmodified Javascript, the fully updated Swift, and several
intermediate versions with individual bugs or groups of bug fixes. These
intermediate versions helped us evaluate the impact of individual or
groups of bug fixes on the algorithm's outputs.

Shadow execution was observational only --- no insulin dosing decisions
were influenced by the experimental code. All data was stored
anonymously. This study was reviewed by the UC Davis Institutional
Review Board and determined to be exempt (IRB ID: 2419522-1).

\subsection{Data-driven replay simulation}\label{data-driven-replay-simulation}

Although mechanistic in silico replay provides end-to-end evaluation, it
doesn't trigger all of the differences we introduced naturally. To
evaluate the full set of divergences, we used a custom data-driven
replay simulation. With this data-driven replay simulation, we start the
simulation in a state known to trigger a bug. Then, we isolate an
individual's underlying physiological and behavioral data using their
real-world glucose and insulin delivery traces. With this data and a
physiological model, we can replay bug inducing algorithm runs and
compare how two or more algorithm variants would perform in closed-loop
using realistic inputs.

Our system builds on established residual signal methods
\cite{patek16, hughes21}. For a given time segment, we compute
\texttt{addedGlucose} as the residual glucose change after subtracting the
estimated effect of active insulin (\autoref{eq:AG}). This residual calculation captures
all unmodeled glycemic inputs \textemdash{} meal absorption, exercise, stress,
sensor noise, and physiological drift -- without requiring explicit
models for all possible inputs.

\begin{equation}
    addedGlucose = \Delta{glucose} + Active_{insulin} \times ISF
    \label{eq:AG}
\end{equation}

For each user, \texttt{addedGlucose} is computed from their recorded CGM and
insulin data using their own programmed insulin sensitivity setting and
the insulin action model from the Swift implementation. During replay,
both algorithm variants receive the identical \texttt{addedGlucose} trace, so
residual errors affect all variants equally. This preserves the validity
of pairwise comparisons between variants even though the simulation does
not attempt to predict absolute glucose trajectories.

In our experiments, we used the shadow execution data to perform
targeted perturbation injection. When shadow execution had detected a
real-world mismatch between algorithm variants, we injected the
divergent system state into the simulator and replayed forward for 24
hours to observe whether the system self-corrected or whether the
divergence cascaded into a clinically significant glucose excursion. 24
hours is the maximum time that a bug can continue to impact results as
this is the furthest lookback time for the Trio algorithm.

To analyze results, we only use the Parkes Error Grid analysis from our
clinical equivalence methodology ($\geq$99\% in Zones A and B). We use this
replay method to provide a realistic set of inputs for the algorithms to
evaluate in a closed loop, not to model virtual humans (see the
Limitations section). Thus, we omit GRI from our analysis.

\section{Results}\label{results}

\subsection{\texorpdfstring{Mechanistic in silico
simulation}{Mechanistic in silico simulation}}\label{mechanistic-in-silico-simulation-1}

\begin{table}[t]
\centering
\begin{tabular}{lcccc}
\toprule
Cohort &
\makecell{Swift\\TIR} &
\makecell{JS\\TIR} &
\makecell{Swift\\GRI} &
\makecell{JS\\GRI} \\
\midrule
Adults       & 93.1\% & 93.0\% & 14.3\% & 14.9\% \\
Adolescents  & 84.1\% & 84.1\% & 21.4\% & 21.2\% \\
Children     & 77.6\% & 77.6\% & 34.9\% & 35.6\% \\
All Patients & 84.9\% & 84.9\% & 23.5\% & 23.9\% \\
\bottomrule
\end{tabular}
\caption{Time in Range and Glycemia Risk Index mean results for patient cohorts on glucose outputs from \texttt{simglucose} running the original Javascript (JS) algorithm and the updated Swift algorithm. The adult cohort age range is between 26 and 68 years old. The adolescent cohort age range is between 14 and 19 years old. The child cohort age range is between 9 and 12 years old.}
\renewcommand{\arraystretch}{1.1}
\label{tab:simulation}
\end{table}

\begin{figure}[t]
    \centering
    \includegraphics[width=\linewidth]{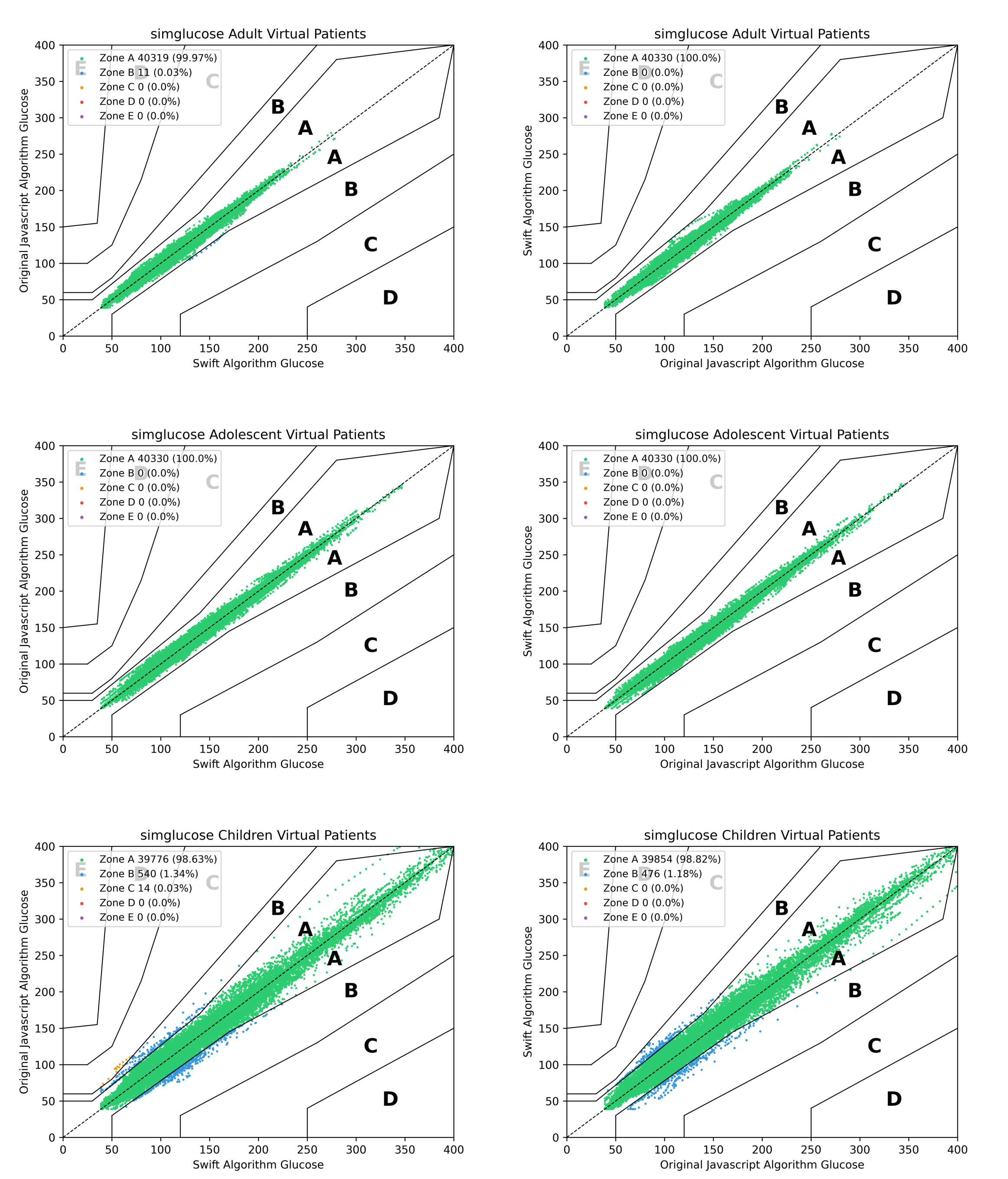}
    \caption{Parkes Error Grids displaying temporal pairwise analysis of glucose outputs from \texttt{simglucose} using the original Javascript algorithm and the updated Swift algorithm to calculate bolus and basal delivery. Each patient cohort has two grids, with the algorithm variant axis assignments reversed between them.}
    \label{fig:figure1}
\end{figure}

We analyzed simulation results from \texttt{simglucose} for 30 virtual patients. Both TIR and GRI were quite similar between the two algorithms with the Swift algorithm performing slightly better (\autoref{tab:simulation}). These metrics suggest the algorithms produce nearly identical outcomes, but this may overlook edge case behavior. In \autoref{fig:figure1}, we use Parkes Error Analysis to evaluate if one algorithm puts the virtual human at increased clinical risk in specific instances.

Parkes Error Analysis assigns points to risk zones based on the
difference between the clinical impact of a test glucose value to a
``ground truth'' reference glucose value. The reference glucose is
plotted on the x-axis, and the test glucose is plotted on the y-axis.
Because there is no strict reference and test differentiation definition
between the two algorithms, we plotted two grids for each patient
cohort, reversing the algorithm axis assignment in the second grid.

\autoref{fig:figure1} shows all adult and adolescent patients were in the lowest risk
area (zone A) for nearly the entire simulation in both versions of the
Parkes Error Grid. 0.03\% of simulation time from adult patients were in
zone B, representing slightly elevated, but clinically insignificant,
risk. Child patients remained within zone A for over 98\% of the
simulation. When the Javascript algorithm was treated as the reference,
child patients spent the remaining 1.18\% of the simulation in zone B.
When the Swift algorithm was treated as the reference, child patients
spent a similar amount of time in zone B (1.34\%). However, child
patients also spent 0.03\% of time in zone C, representing significant
glucose differences.

Overall clinical equivalence was maintained between the two algorithms
with minimal differences between TIR and GRI (\autoref{tab:simulation}). The difference
in GRI stayed well below our clinical equivalence threshold of $\Delta GRI \leq
10$. In a Parkes Error Analysis, virtual patients spent at least 99\% of
the simulation in low risk zones A and B (\autoref{fig:figure1}). While rare,
clinically significant differences did occur between the \texttt{oref}
algorithms. Our experiments revealed floating point precision bugs in
the Javascript algorithm were responsible for this divergent behavior.
See the Discussion section for an explanation of how precision
differences impacted the deviations.

\subsection{Shadow execution}\label{shadow-execution-1}

We analyzed 736,480 algorithm invocations across eight Trio users. \autoref{tab:shadow_mismatches} summarizes the mismatch rates between the original Javascript and
updated Swift implementations. All mismatches exceeded the approximate
matching thresholds defined by Trio \cite{triotest}.

\begin{table}[t]
\centering
\begin{tabular}{lcc}
\toprule
Function & Invocations & Mismatches \\
\midrule
\texttt{iob}             & 328,146 & 1,420 (0.43\%) \\
\texttt{autosens}        & 29,527  & 360 (1.22\%) \\
\texttt{determineBasal}  & 184,765 & 134 (0.07\%) \\
\texttt{meal}            & 194,042 & 17 (0.01\%) \\
\bottomrule
\end{tabular}
\caption{Shadow execution mismatch counts between the original Javascript algorithm and the updated Swift implementation for the four main \texttt{oref} functions.}
\renewcommand{\arraystretch}{1.1}
\label{tab:shadow_mismatches}
\end{table}

We tested eight \texttt{iob} bug fixes individually, and one dominated: the
\texttt{splitTimespan} error \cite{king26}. This error incorrectly dropped pump
events from the IOB accounting, and the fix accounted for 99.7\% of \texttt{iob}
differences and 100\% of \texttt{meal} differences. For \texttt{autosens}, the same fix
drove 73\% of differences, while a hard-coded 8-hour suspend duration
\cite{king26} produced the largest individual deltas (max 0.50).

Of the 1,420 \texttt{iob} mismatches, 95 (0.03\% of all invocations) flipped the
sign of Net IOB \cite{lal21, riddell26}, a clinically meaningful
threshold. Sign changes for Net IOB change how users interpret whether
they have excess or deficit insulin on board, and negative Net IOB
values will cause the algorithm to backfill insulin, potentially leading
to overdosing. Four \texttt{autosens} mismatches (0.01\%) flipped the sensitivity
classification between resistant and sensitive.

Although our results showed that our fixes did introduce clinically
meaningful differences at the level of individual algorithm invocations,
the question remains about whether these differences translate to
differences in overall glycemic control.

\subsection{Data-driven replay simulation}\label{data-driven-replay-simulation-1}

From the shadow execution log, we selected the top 100 bug invocations
for both \texttt{iob} and \texttt{autosens} functions, ordered by the delta between the
original Javascript and updated Swift. Output differences in \texttt{iob} range
from 0.541 to 0.332, while \texttt{autosens} outputs range from 0.5 to 0.07.
Following each trigger, we simulate 24 hours of added glucose traces to
look at the downstream effect of the bug/difference. Each bug invocation
was tested with 20 added glucose traces sampled from the trio user
group, to simulate for each bug invocation. This methodology yielded
558,144 comparative pairs for \texttt{iob}, 552,672 comparative pairs for
\texttt{autosens}.

\begin{figure}[t]
    \centering
    \includegraphics[width=\linewidth]{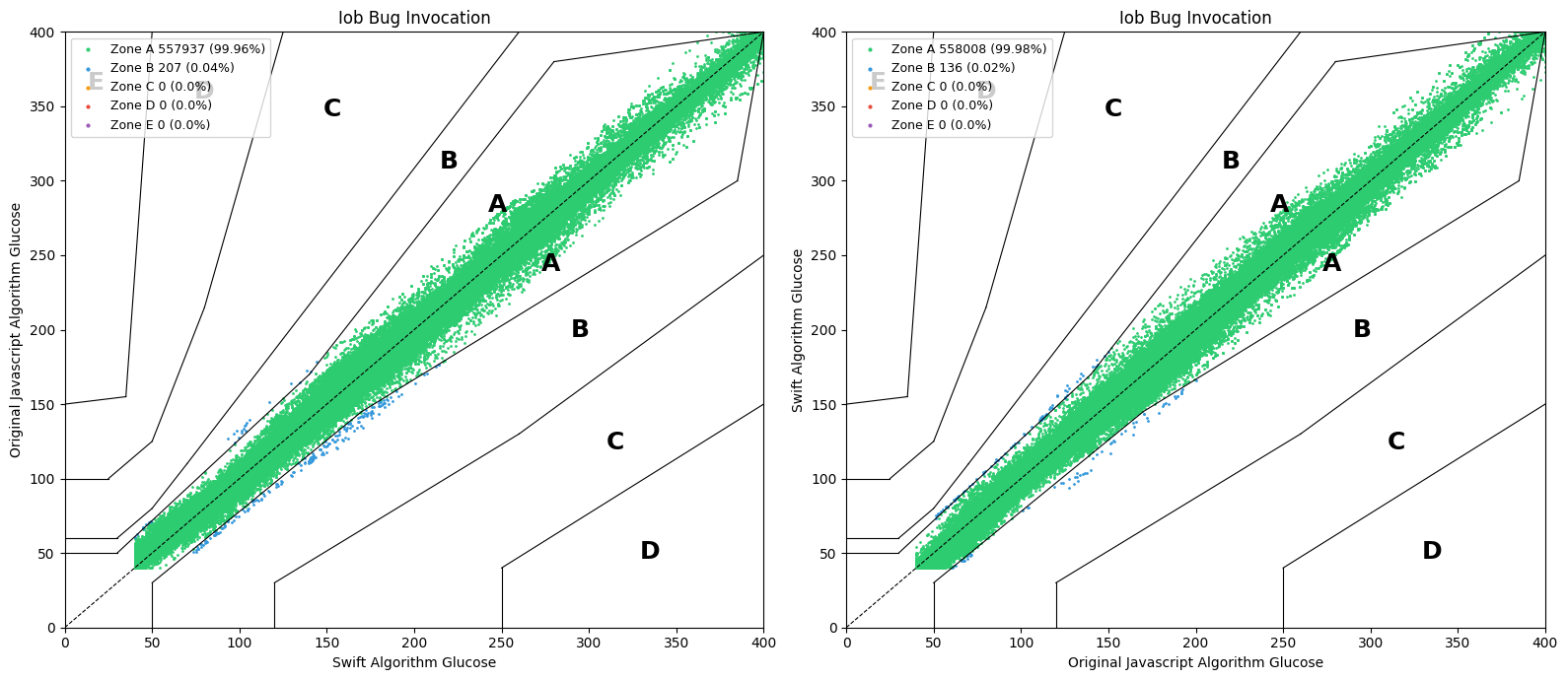}
    \caption{Parkes Error Grid displaying pairwise analysis of glucose outputs from replay simulation  of \texttt{iob} bugs.}
    \label{fig:figure2}
\end{figure}

\begin{figure}[t]
    \centering
    \includegraphics[width=\linewidth]{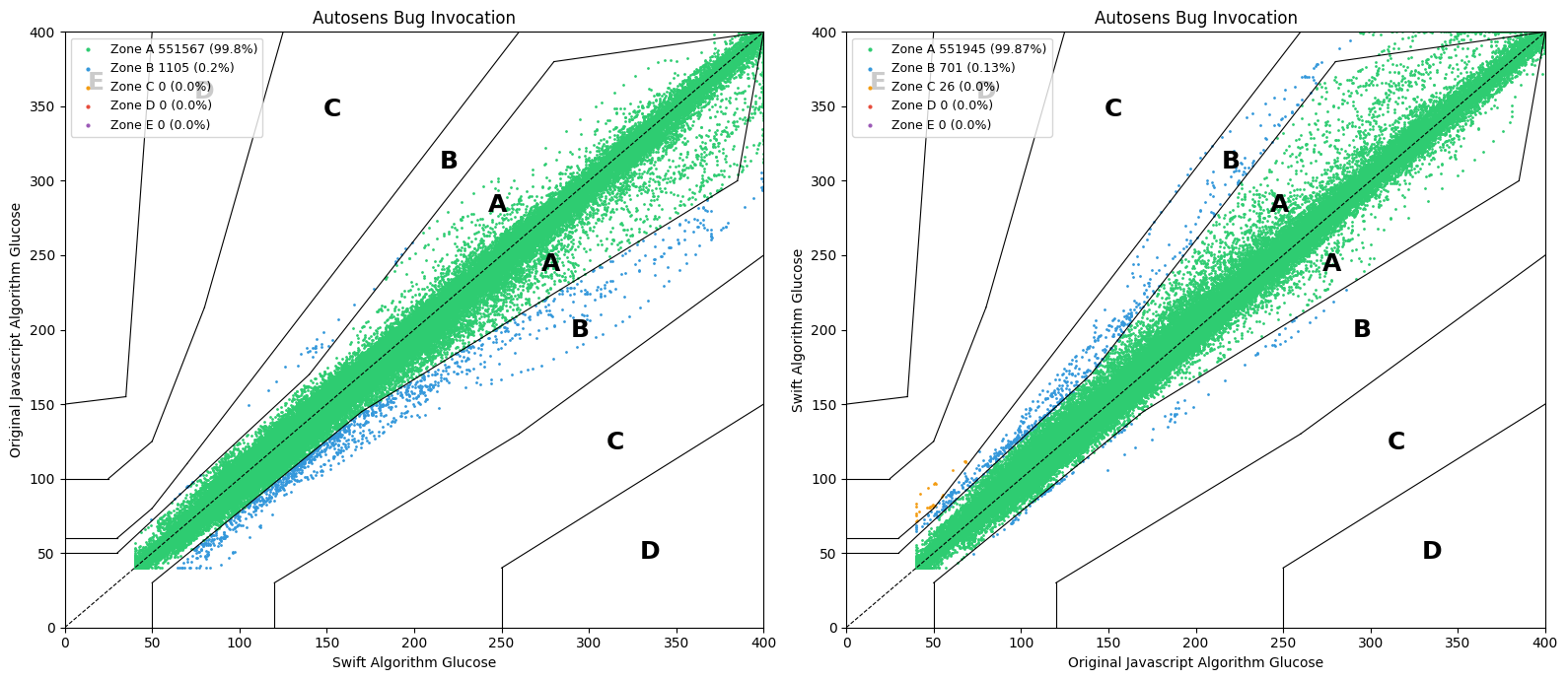}
    \caption{Parkes Error Grid displaying pairwise analysis of glucose outputs from replay simulation  of \texttt{autosens} bugs.}
    \label{fig:figure3}
\end{figure}

Both Parkes Error Grids (\autoref{fig:figure2} \& \autoref{fig:figure3}) showed that the downstream effects when the Swift implementation deviates from the Javascript
implementation remain within our clinical equivalence threshold of $\geq$99\%
of pairwise glucose values in Parkes Zones A and B.

\section{Discussion}\label{discussion}

In this section, we discuss three case studies where there was a
clinically meaningful difference between the simulated outcome for the
original Javascript algorithm and the updated Swift implementation.
Although our results showed that the Swift algorithm is clinically
equivalent to the Javascript implementation (see Results),
these case studies provide insight into the nature of AID algorithms and
illuminate failure modes that generalize beyond this port, suggesting
design principles for more robust AID algorithms.

\subsection{Case study: Discontinuity in UAM forecasting amplifies a small IOB error}\label{case-study-discontinuity-in-uam-forecasting-amplifies-a-small-iob-error}

\begin{figure}[t]
    \centering
    \includegraphics[width=\linewidth]{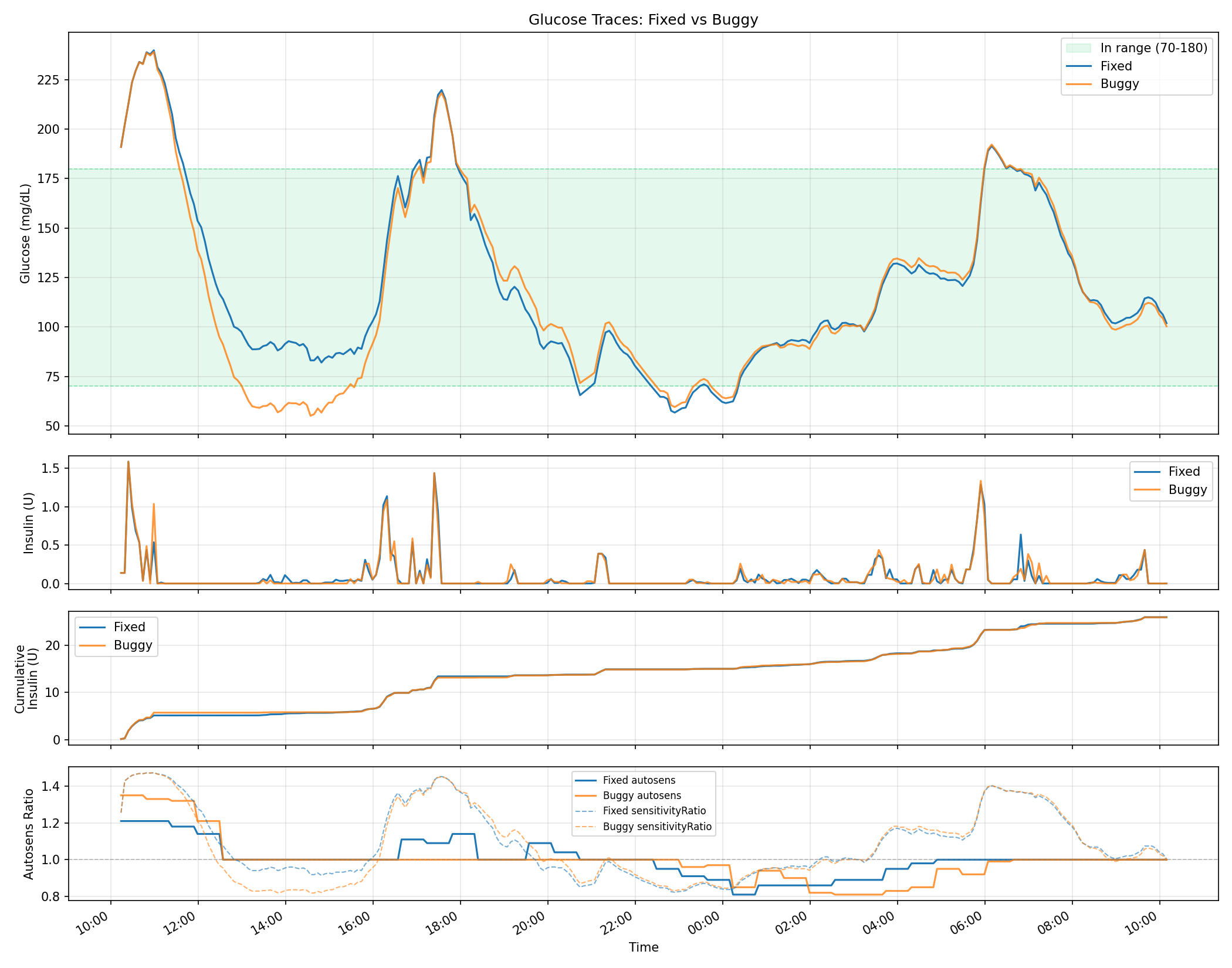}
    \caption{24 hour glucose, insulin delivery, autosens, and dynamic ISF values for both the Swift and Javascript algorithms to simulate the effects of an IOB bug.}
    \label{fig:figure4}
\end{figure}

This case study illustrates the cross-component cascade our framework
anticipates: a factual bug in IOB whose impact only becomes visible
after propagating through two heuristic components downstream. The bug
is the ``Hard-coded DIA of 8 hours'' issue listed in the Trio GitHub
repository \cite{king26}. \autoref{fig:figure4} shows the 24-hour glucose, insulin
delivery, \texttt{autosens}, and \texttt{DynamicISF} traces for both algorithms.

The chain has five hops. An IOB bug perturbs the historical insulin
values that \texttt{autosens} consumes, so the two algorithms compute different
\texttt{autosens} ratios from identical pump histories (1.35 in Javascript versus
1.21 in Swift). The simulated user runs \texttt{DynamicISF} with the ``sigmoid''
function, which adjusts dosing sensitivity directly from glucose rather
than from \texttt{autosens}. However, the \texttt{autosens} ratio still rescales basal
contributions inside the Net IOB calculation itself, producing a small
but persistent offset in active insulin at each historical timestep,
even though \texttt{DynamicISF} is nominally in charge of
sensitivity.

This small offset propagates into the Unannounced Meal (UAM) forecasting
algorithm, which selects the earliest local minimum in its \texttt{deviation}
series as a slope anchor for projecting carbohydrate impact. Here the
discontinuity fires: a sub--1 mg/dL difference in a single \texttt{deviation}
value is enough to flip which bucket each algorithm selects. The Swift
algorithm latches onto a later bucket with steeper decay, projecting
carbohydrate impact to resolve in 75 minutes (forecast: 140 mg/dL). The
Javascript algorithm latches onto an earlier bucket with shallower
decay, projecting continued digestion for 170 minutes (forecast: 223
mg/dL). \texttt{determineBasal} reads the resulting 83 mg/dL forecast gap as an
unmet insulin requirement and issues a larger super-microbolus into a
\texttt{meal} that is already resolving.

\textbf{Takeaway:} When an algorithm includes discontinuous outputs from
continuous inputs, small differences in values, like we observed with
this IOB bug, can lead to large differences in dosing decisions. This
effect is also why our framework classifies bugs by where the error
originates rather than where its effects appear.

\subsection{Case study: Factual bug silently cascaded to dosing decision}\label{case-study-factual-bug-silently-cascaded-to-dosing-decision}

\begin{figure}[t]
    \centering
    \includegraphics[width=\linewidth]{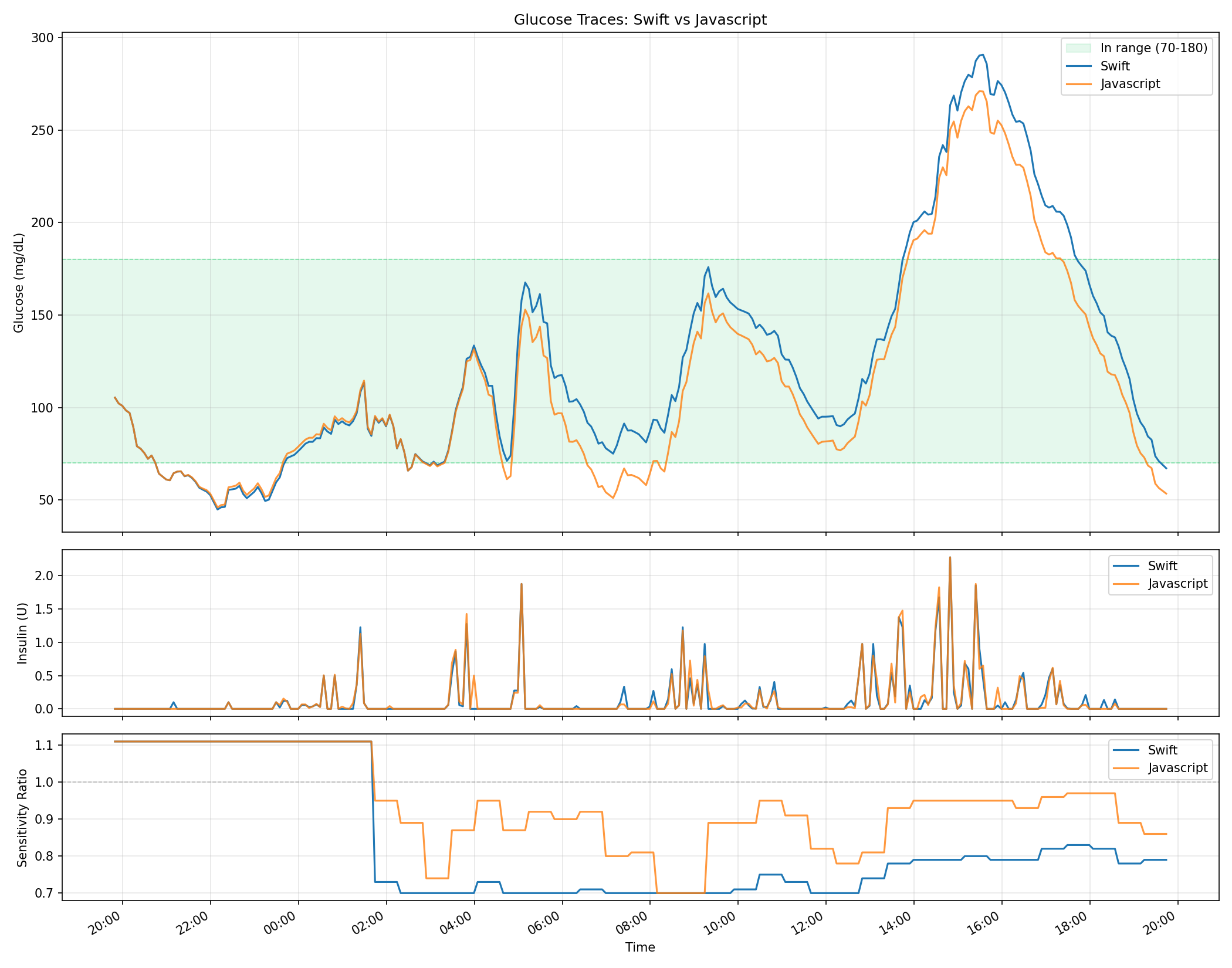}
    \caption{24 hour glucose, insulin delivery, and autosens values for both the Swift and Javascript algorithms to simulate the effects of the bug. Both Javascript and Swift use initial autosens ratios from the captured inputs at the beginning until there are 8 hours of glucose records to allow the recalculation of autosens.}
    \label{fig:figure5}
\end{figure}

Where the previous case study showed how a small IOB perturbation could
be amplified by a discontinuity, this case study shows the opposite
failure mode: a factual bug that produces no dramatic amplification,
just a slow accumulation of bias that the system has no way to detect.
The bug is the ``\texttt{splitTimespan} logic can drop pump events'' issue listed
in the Trio GitHub repository \cite{king26}. \autoref{fig:figure5} shows the
resulting 24-hour glucose, insulin delivery, and \texttt{autosens} traces for
both algorithms.

The bug silently drops temp-basal segments from pump record accounting.
When a record is dropped, the algorithm has no record of that interval
and falls back to the scheduled basal rate, logging insulin delivery
that did not actually occur. In this instance, the dropped record was a
zero temp-basal command --- meaning the algorithm had suspended basal
entirely --- so the Javascript algorithm believed scheduled basal had
been delivered when in fact none had.

The error then accumulates through \texttt{autosens}. Both algorithms hold their
initial \texttt{autosens} ratios for the first eight hours until enough glucose
history exists to trigger a recalculation. At that first recalculation,
the bias becomes visible: 0.95 for Javascript, 0.73 for Swift. Because
Swift accounts for the dropped record correctly, it infers the
individual is more insulin-sensitive than Javascript does, and the
divergence compounds across subsequent \texttt{autosens} steps. Javascript
ultimately issues an extra +0.5 U bolus that Swift correctly omits.

\textbf{Takeaway:} Factual bugs must be prioritized; because they
corrupt the system's objective source of truth, these errors silently
cascade to downstream heuristics and dosing decisions without triggering
system crashes or user warnings.

\subsection{Floating point imprecision bug}\label{floating-point-imprecision-bug}

While conducting mechanistic in silico simulations (\texttt{simglucose}),
floating point representation differences between the two \texttt{oref}
algorithms were the main cause for divergent outputs. There is a
fundamental difference between how the two algorithms handle floating
point math. The Javascript algorithm uses 64-bit floating point numbers,
which can create small numerical imprecisions during mathematical
operations. In native Javascript, for example, 0.1 + 0.2 equals
0.30000000000000004 rather than 0.3. In contrast, the Swift
implementation uses exact decimal math to avoid precision errors.

The \texttt{oref} algorithm predicts the user's future physiology using
forecasting variables computed from historical time series data, such as
IOB, to influence the temp basal rate and duration calculation. At any
timestep, mishandling floating point precision can diverge the
algorithm's internal state from the correct value. The algorithm never
corrects the error. At each subsequent timestep, more time series data
is computed using the incorrect history. Over time, the small errors in
the previous time series data can collectively accumulate enough of a
difference that forecasting variables are rounded to an incorrect value.

\textbf{Takeaway:} Rounding and floating point precision semantics are
not just cosmetic implementation details. Floating point imprecision has
real unintended impacts on \texttt{oref} algorithm commands. For software that
impacts insulin delivery decisions, all mathematical operations should
be done using precise math to avoid propagating rounding errors into
future commands. Floating point math should only be used when
performance time is critical and errors can be absorbed without
propagating it forward into critical decision-making functions.

\subsection{Framework Applicability Across OS-AID
Systems}\label{framework-applicability-across-os-aid-systems}

In \nameref{results}, our shadow execution experiments comparing the original
Javascript implementation and Swift port of Trio's \texttt{oref} algorithm
revealed eight \texttt{iob} bugs in the Javascript implementation. These bugs
were responsible for mismatches in outputs between the two algorithms
(\autoref{tab:shadow_mismatches}). We analyzed whether other OS-AID systems, specifically
OpenAPS, iAPS, AndroidAPS, and Loop, contain equivalent or similar bugs.

OpenAPS and iAPS execute the same \texttt{iob} algorithm as Trio. In OpenAPS, all
eight bugs are present, and in iAPS, seven of the eight bugs are
present. AndroidAPS and Loop run different \texttt{iob} algorithms then Trio,
meaning we cannot verify if these eight \texttt{iob} bugs exist in these systems.
While we do not have specific bugs proving this framework's usability
for these systems, it is highly probable their algorithms contain
similar bugs to the ones we found in Trio. Manual code inspection of the
AndroidAPS and Loop codebases revealed these two systems are heuristic
based and do not use precise math; the exact conditions which trigger
bugs in Trio. These findings suggest our framework, which evaluates the
clinical impact of bugs and guides the decision of whether or not to
patch them, can be broadly applied to various OS-AID systems.

\subsection{Limitations}\label{limitations}

Our study has several important limitations regarding simulation
fidelity and cohort diversity. First, while our data-driven replay
simulation accurately preserves real-world sensor noise and unmodeled
inputs by building on established residual signal methods, it is
inherently constrained by a linear assumption of insulin action. If the
algorithmic divergences between the Swift and Javascript implementations
lead to drastically different insulin dosing trajectories, the
simulation cannot reliably predict complex, non-linear physiological
counter-regulatory responses (e.g., endogenous glucose production during
severe hypoglycemia). Therefore, our 24-hour replay analysis provides
rigorous safety bounds for expected algorithmic deviations, but it does
not simulate catastrophic algorithmic failures.

Second, while the \texttt{simglucose} environment provides a rigorous,
mathematically validated physiological model for stressing algorithm
control logic based on the 2007 Dalla Man equations \cite{dallaman07},
it is an open-source implementation rather than the proprietary,
FDA-accepted commercial UVA/Padova simulator. Additionally, \texttt{simglucose}
is not a comprehensive real-world representation of an individual with
diabetes. It is focused on modeling the metabolism of, and interaction
between, glucose and insulin; it does not include all confounding
factors from the real-world known to influence glucose levels, such as
hydration level or illness. While this simulator is useful to help
evaluate the effect software patches have on algorithmic outputs, it
does not guarantee a one to one correspondence to real everyday use.
Therefore, our mechanistic in silico simulation results are
intended to demonstrate clinical equivalence and the functional
integrity of our decision framework, rather than to make absolute
predictions of clinical efficacy for regulatory purposes.

\subsection{Conclusions}\label{conclusions}

Applied to the Trio \texttt{oref} Swift port, the framework confirmed that
patched implementations can achieve clinical equivalence to their
validated predecessors in simulation. Notably, the original Javascript
algorithm performed well even in the presence of these bugs with high
TIR and low GRI results in \texttt{simglucose} simulations, reflecting the
robustness of the algorithm. We also showed why fixing factual and
computational bugs are important: those bugs can accumulate and cascade
through algorithmic discontinuities into divergent insulin dosing
decisions.

\bibliographystyle{plainnat}
\bibliography{bug_finding}

\end{document}